\documentclass[doublespace,11pt]{article}
\usepackage{graphicx}
\usepackage{setspace}
\usepackage{subfigure}
\usepackage{epsfig}

\usepackage{url}

\begin{document}
\sloppy
\begin{spacing}{1}

\begin{titlepage}
\hspace{0.08in}
\begin{minipage}{\textwidth}
\begin{center}
\vspace*{3cm}
\begin{tabular}{c c c}
\hline
 & & \\
 & {\Huge The University of Algarve} & \\
 & & \\
 & {\Huge Informatics Laboratory} & \\
 & & \\
\hline
\end{tabular}\\
\vspace*{2cm}
{\Large
UALG-ILAB\\
Technical Report No. 200701\\
January, 2007\\
}
\vspace*{3cm}
{\bf A Virtual Logo Keyboard for People with Motor Disabilities}\\

\addvspace{0.5in}
{\bf St\'{e}phane Norte}, and {\bf Fernando G. Lobo}\\
\vspace*{-0.1in}
\vspace*{4cm}
Department of Electronics and Informatics Engineering\\
Faculty of Science and Technology \\
University of Algarve \\
Campus de Gambelas\\
8000-117 Faro, Portugal\\
URL: {\verb http://www.ilab.ualg.pt }\\
Phone: (+351) 289-800900\\
Fax: (+351) +351 289 800 002 \\
\end{center}
\end{minipage}
\end{titlepage}

\title{\bf A Virtual Logo Keyboard for People with Motor Disabilities}
\author{    {\bf St\'{e}phane Norte}\\
            \small UAlg Informatics Lab\\
            \small DEEI-FCT, University of Algarve\\
            \small Campus de Gambelas\\
            \small 8000-117 Faro, Portugal\\
            \small snorte@ualg.pt
\and
            {\bf Fernando G. Lobo}\\
            \small UAlg Informatics Lab\\
            \small DEEI-FCT, University of Algarve\\
            \small Campus de Gambelas\\
            \small 8000-117 Faro, Portugal\\
            \small flobo@ualg.pt
}
\date{}
\maketitle

\begin{abstract}
In our society, people with motor impairments are oftentimes socially
excluded from their environment.
This is unfortunate because every human being should have the possibility
to obtain the necessary conditions
to live a normal life.
Although there is technology to assist people with motor impairments, few systems
are targeted for
programming environments.
We have created a system, called Logo Keyboard, to assist
people with motor disabilities to program with the Logo programming language.
With this special keyboard we can help more people to get involved into computer
programming and to develop projects in different areas.
\end{abstract}

\section{Introduction}
Computer programming is an important activity occupying a large number
of students in our society. Disabled people have reduced opportunities
in a lot of areas, and the field of education
is no exception.
A virtual environment offers the possibility to control or operate in
the real world partly alleviating physical limitations.
Most people use their arms, hands, and fingers when using a computer,
namely a keyboard and/or mouse.
Nevertheless, for people with motor disabilities, the need for controlled
movements can be an obstacle
for them to be able to interact with the computer.
There are many types of physical disabilities that affect a pleasant
human-computer interaction. Several
diseases cause muscle deterioration and/or neurological disorders
that can result in physical weakness, loss of
muscle control, paralysis, and amputation. The main problem for people
with physical disabilities is
the capacity to access computer controls (i.e. power switch, disk drives,
monitor adjustments) and the ability to
type on the standard keyboard or move pointing devices like a mouse.

People with motor impairments can use alternative keyboards which replace
the standard keyboard for
input of information into the computer.
In this paper, we present the virtual Logo Keyboard. With this tool
we provide an easy way to help people program in Logo.
Our work focuses on Logo because it is an excellent programming environment.
The Logo Programming language was designed as a tool for learning.
It is not only extensible, but flexible and interactive.
For most people, learning Logo is not an end in itself, but a beginning.
The Logo programming language is useful to create
activities in science, language, music, robotics, telecommunications and
mathematics~\cite{Resnick:09}.
MicroWorlds, StarLogo, UCBLogo, and MSWLogo, are examples of some of the
most popular Logo interpreters.

One of the most significant motivations for including Logo in the
beginning of educational computing is that
Logo presents an example of a way in which computers can be used
to reconceptualize the teaching and learning
process, rather than simply to improve or enhance traditional
forms of learning. From the beginning, Logo was
created to be a mathematical learning environment, as well as a
programming language~\cite{Hoyles:12}.
Seymour Papert, Logo's principal creator, based his inventions on
his years of work with Jean Piaget, the Swiss
philosopher and psychologist who has done so much to promote
understanding of the ways in which children construct
their own learning in natural settings~\cite{Piaget:13}.
Using Logo as a programming language is also an excellent introduction
into the world of computer
science and programming. Programming is a key intellectual activity
associated with both the many uses of computers
in the real world, and the educational uses of computers in elementary
and middle schools and beyond.

In the next section, we present some background related to this work.
Then, section 3 presents the development of the virtual Logo Keyboard
and its motivation.
Section 4 describes usability tests conducted with 3 children.
Finally, we describe an outline of future work in
section 5, and the conclusions in section 6.

\section{Background}

There is a large amount of work in assistive technology and in building
computers more accessible to people
with disabilities and learning difficulties~\cite{Trewin:10,Demasco:11}.
However, there is less work focused
on assistive technology to help people to program in an educational environment.
This section describes previous research on virtual keyboard technology
specifically designed to provide
people with motor impairments the ability to interact with a computer.

\subsection{Virtual keyboard technology}
Virtual tools can facilitate the manipulation of computer devices~\cite{Zhai:18}.
Most virtual keyboards are based on the physical layout of a standard keyboard~\cite{Brad:08}.
Before the user can select anything on the screen, she need physical instruments for
a direct interaction~\cite{Keates:15}.
Many devices are used to obtain this access. For example, switches, pointing devices,
joysticks, physical keyboards, mouses, trackballs, and webcams.
Some people may have difficulties in using a standard keyboard.
The Logo Keyboard uses a scanning system to help people interact with the computer.
The scanning method is widely used among the disabled
in various computer-aided systems. Scanning techniques allow
the user to use one switch
to change the focus from one item to another.
The focus automatically shifts from item to item after a predefined time has elapsed
and all the user has to do is to press a single switch or key
when the desired item is in focus~\cite{Steriadis:17,Majaranta:16}.

\section{Development of the Virtual Logo Keyboard}
The Logo Keyboard was developed with the purpose of aiding and
integrating people with motor impairments
in a programming environment. Some studies reveal that people
with certain limitations, motor or
cognitive, can develop programs~\cite{Valente:03}. Unfortunately,
many children and adolescents remain
frustrated in schools due to a lack of  assistive devices.
Through this interface we make it possible for users, or carriers
of impairments to use the Logo language.
The Logo Keyboard can be used with a Logo interpreter, for example,
MicroWorlds Pro, UCBLogo, or MSWLogo.
The interaction of the user with the virtual Logo Keyboard begins
with a touch on an assistive device (mouse,
switch, or keyboard). The message is sent to the Logo Keyboard and the
chosen Logo command is sent to
the command window (output stream) where it is executed by the
Logo interpreter.

\subsection{Logo Keyboard Interface}
\begin{figure}
\begin{center}
\includegraphics{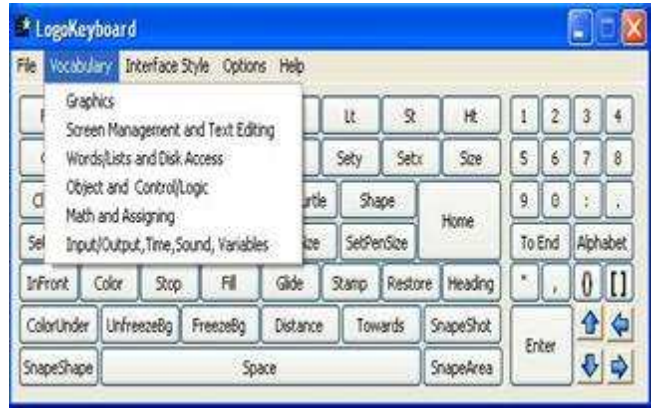}
\end{center}
\caption{Logo Keyboard Interface - Vocabulary Menu.}
\label{fig:logok11}
\end{figure}

The Logo Keyboard interface was created in C++. It contains an adjusted
environment to support different kinds of aids.
The keyboard was created with special keys to produce an output of Logo
programming commands.
Each vocabulary was put into separate groups allowing the creation of
an organized form of a keyboard
with 312 keys (Figure~\ref{fig:logok11}).

\begin{figure}
\begin{center}
\includegraphics{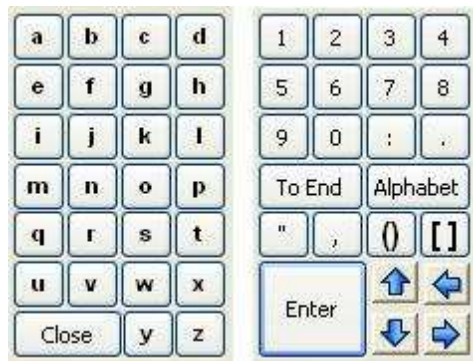}
\end{center}
\caption{Alphabetic and numeric keyboard.}
\label{fig:alph}
\end{figure}

In addition to the Logo programming commands, there's the possibility
of using a numeric or alphabetic
keyboard (see Figure~\ref{fig:alph}).

\begin{figure}
\begin{center}
\includegraphics{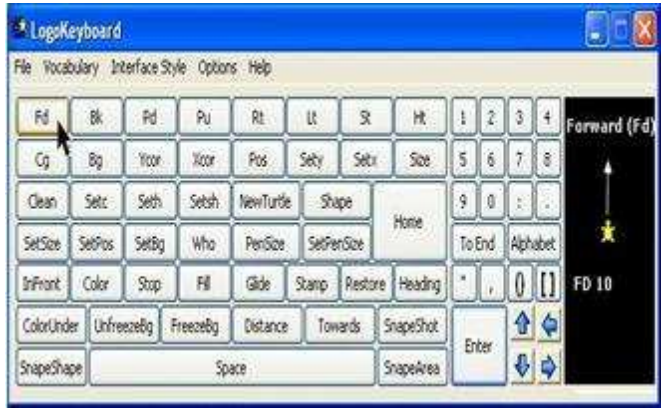}
\end{center}
\caption{The user places the mouse under the intended key and
visualizes the help system about that key / command.}
\label{fig:help}
\end{figure}

The Logo Keyboard has a system that allows the user to modify
the interface transparency depending
on the needs of the users.
The options menu have useful features. It includes an
integrated help system providing information about each command
along with a brief description and an animated example of its use
(see Figure~\ref{fig:help}).

\begin{figure}
\begin{center}
\includegraphics{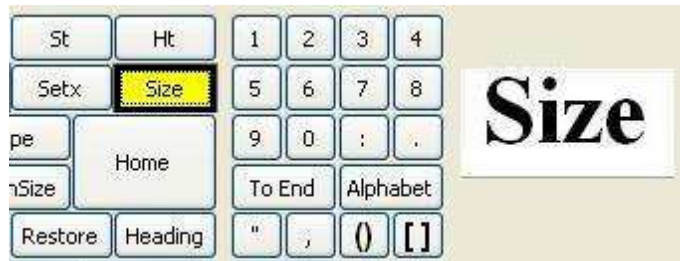}
\end{center}
\caption{The scanning system allows zooming a command without the use of a mouse.}
\label{fig:zoom1}
\end{figure}

The zoom options help people with visual impairments to zoom any given key.
This feature is extremely useful for people with low vision. It can be used with
the assistance of the mouse, placing it in the intended key
or automatically through the scanning system (see Figure~\ref{fig:zoom1}).

The user can configure several options of the keyboard, such as:
scanning velocity, number of repeat scanning cycles,
scanning sound, scanning color, and keyboard size.
The user also has the possibility to identify Logo commands with
a voice system when the mouse moves over them.

One of the aspects that was carefully thought about was the
possibility to develop an easily accessible
and configurable interface. A person with disabilities
can press a single button to configure or use the
interface without assistant's help (except if the option
of the aid commands is turned on).
Through the scanning mechanism, the keyboard can completely
assist a user with motor impairment~\cite{Condado:07}.

\subsection{Input Methods}
\begin{figure}
\begin{center}
\includegraphics{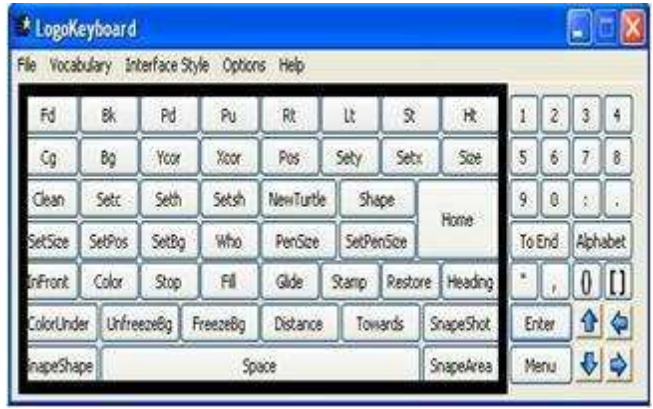}
\end{center}
\caption{An example of the scanning system. There are 2
principal groups of keys (Logo commands and
numeric/alphabetic), and the cursor advances through
each group at a specified rate.}
\label{fig:scan}
\end{figure}

\begin{figure}
\begin{center}
\includegraphics{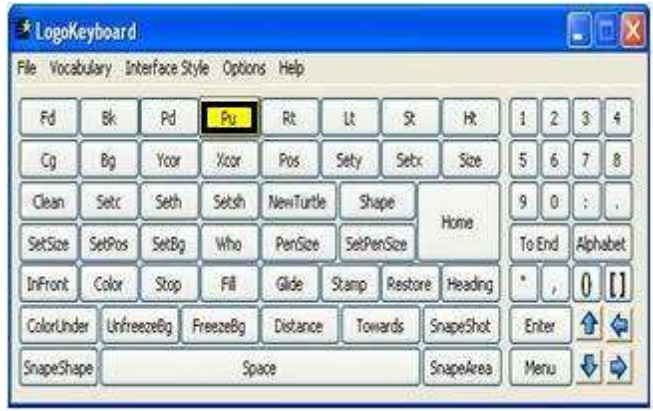}
\end{center}
\caption{In this example, the first row of the first
group was selected, and now, the scanning system
becomes focused on the keys of that subgroup, with
the cursor advancing through each of the Logo commands one at a time.}
\label{fig:rowscan}
\end{figure}

People who are able to use a mouse, can easily use the Logo Keyboard to build Logo programs
using a simple click of the mouse under the desired key. It is important to pay attention to
where the last mouse focus was. The principal target of the Logo Keyboard is to be used with a
Logo interpreter, but it can also be used to write some text with the alphanumeric keyboard.
People who cannot use a mouse are able to use the scanning mechanism of the keyboard.
There are two important scanning groups (Figure~\ref{fig:scan}):
The first is used to select the programming keyboard keys;
The second is used to select items in the numeric or alphabetic keyboard.
After the user has selected one of the groups, the selection is recognized
and the scanning proceeds under subgroups of the original group.
Eventually, the scanning system focuses on single keys, one at a time
(see Figure~\ref{fig:rowscan}).
In this phase the next click will be the choice of the desired key.
To use the scanning system, the user
needs to press the ``Space'' key or use a switch.
The Logo Keyboard presents important features, facilitating and increasing
the writing speed of issuing commands.

\subsection{Motivation and Learning with Logo keyboard with disabled Children}
The experience with Logo philosophy using computers with children
with disabilities shows that computers
can offer a new way to educate~\cite{Goldenberg:01}.
The computer can be a tool with which the child can develop
abstract thought, express ideas previously inaccessible,
and be an active and creative element.
Logo has a lot of characteristics to help create an ideal
learning environment for children with disabilities.
Primarily, the materials that the child has to operate
in the Logo environment are not physical objects that require
a high degree of motor coordination. The objects are controlled
by the computer. The Logo Keyboard is an instrument
that helps to minimize the barriers between the child
and the physical world. Thus, if the child has enough motor
coordination to control a single touch button, she can control
the Logo Keyboard and do everything that they want
without needing aid of other people.
The activities that the child develops in a Logo programming
environment can be determined by  their own interest
and imagination. This is very important because it allows
the keyboard to transform the passivity of the child with
disabilities. The Logo exercise allows the formation of the
intuitive knowledge of the child and it allows others to
see the process that it uses to develop a specific activity.
The perspective of education in accordance with the
Logo philosophy consists of giving children the power to
reach their potential and to forget about their disabilities.
As advocated by Seymour Papert, children should be provided with
a proper environment for being constructors of their
own knowledge, learning along the way with their own mistakes.
The technology will help us to provide more opportunity for
teachers to work with students on
projects where they will mutually implement their ideas ~\cite{Papert:05,Papert:06}.

\section{Usability tests}
We have made usability tests of the Logo Keyboard with students of a highschool.
The students learned the introduction to Logo programming with a regular keyboard,
but we soon realized than some of them revealed physical motor impairments to
control a regular keyboard.
To help these children we used the Logo Keyboard and we created some experiments.
The population
of the study consisted of three students between the ages of 15 and 17.
The students who have been investigated
have different characteristics. One of them does not have any motor
impairment although the others do have.
Each student was assigned to write
the same program (shown in Figure~\ref{fig:program}), using
different techniques to make it. The users with disabilities
used the Logo Keyboard
and the others wrote the program with a regular keyboard.
Table~\ref{tab:table1}
summarizes the total number of key-presses made and the time taken.

\begin{figure}
\begin{center}
\begin{verbatim}
                             to square
                               make "n 4
                               repeat (:n) [fd 30 rt 90]
                             end
\end{verbatim}
\end{center}
\caption{Program used in the experiment.}
\label{fig:program}
\end{figure}

\subsubsection*{Daniel}
Daniel is a 15 year old boy without motor impairments.
To create projects in Logo he uses a physical keyboard at school.
When given the opportunity to learn the Logo Keyboard in the classroom, Daniel used one of
the existing options of the keyboard. The feature used was the help system about the
Logo commands.
He said: ``This feature is excellent to learn the Logo language and how to use it''.
Daniel confirmed that with this tool, any pupil could learn by himself how to program in Logo.
His evolution was fantastic; he constructed his own knowledge with his own mistakes.

\subsubsection*{Ana}
Ana is a 16 year old girl with motor impairments. She needs the aid of a wheelchair to be able to move.
Her interaction with computers is good but she does not like to use a conventional mouse.
Ana prefers using a trackball because it is less tiring for her. To program in Logo,
she prefers  using the Logo Keyboard scanning system. In spite of a slightly longer delay,
she liked the experiment. Ana was able to reach each goal and she admitted that at home
she wanted to use a trackball to continue developing Logo projects with the help of
the Logo Keyboard.

\subsubsection*{Nilton}
Nilton is a 17 year old boy, with poorly developed fingers. Each hand has three fingers
of extreme size. This disability stops him from exercising diverse tasks at school.
In the classroom, Nilton learned how to program in Logo, but we quickly understood that
he was frustrated due to the fact that he could not keep up with his colleagues.
Nilton used the Logo Keyboard to continue his work. Using a mouse to control the
interface, Nilton was able to program very quickly and to exceed some of his colleagues.
Throughout the school year, Nilton's confidence grew and he exceeded his limitations
showing that he was a good programmer. A direct quote from him follows:
``With the Logo Keyboard I had the possibility to enjoy the same rights as my
colleagues - Learn Logo''.

\begin{table}
\caption{Summary of the typing tests.}
\label{tab:table1}
\begin{center}
\begin{tabular}{|c|c|c|c|}
\hline
Student & Device & No. of Key-presses      & Time (secs)\\
\hline \hline
  Daniel  & Physical Keyb.  & 56        & 46     \\ \hline
  Ana     & Virtual Keyb.   & 108       & 404    \\   \hline
  Nilton  & Virtual Keyb.   & 27        & 73     \\ \hline
\end{tabular}
\end{center}
\end{table}

\section*{Results}
At the end of the experiment, a small questionnaire was given to the students
to understand the degree of satisfaction
concerning speed, usability and precision of the Logo Keyboard (table~\ref{tab:table2}):

\begin{table}
\caption{Subjective ratings of speed, efficacy and precision when using the Logo
Keyboard and a physical keyboard.
Ratings were provided on a scale from 1 (weak) to 5 (good).(P-Physical V-Virtual)}
\label{tab:table2}
\begin{center}
\begin{tabular}{|c|c|c|c|c|c|c|}
\hline
& \multicolumn{2}{|c|}{Speed} &  \multicolumn{2}{|c|}{Usability} & \multicolumn{2}{|c|}{Precision}\\
  Student / Keyboard& P & V  & P & V & P & V\\
\hline \hline
  Daniel  &    5    &    4     &    4    &    4     &    4    &     5    \\ \hline
  Ana     &    1    &    2     &    2    &    5     &    2    &     4     \\   \hline
  Nilton  &    2    &    5     &    2    &    5     &    3    &     4     \\ \hline
\end{tabular}
\end{center}
\end{table}

\section*{Discussion}

It was verified that the Logo Keyboard was able to assist the students
positively by increasing their performance,
speed and accessibility to program.
From the tests we can observe that an individual that
can manipulate a mouse to use the Logo Keyboard
is able to program more easily than an individual that uses a conventional keyboard.

It is verified that the use of the scanning system can generate a
longer delay to select a desired Logo command,
but the amount of characters does not become a negative aspect, due
to the fact that a complete Logo command can be selected at once, rather
than typing the individual characters that constitute the command.
This can potentially improve their overall typing rate.
The users who do not have motor impairments do not have great advantages
in the use of the Logo Keyboard other than using the integrated help system.
The main goal of the sample program used in this study was to test the
biggest difficulties felt by the students with disabilities---the simultaneous
manipulation of two keys which, on a regular keyboard, are required to produce
special symbols that often appear in computer programs
(for example, {\tt [, ], :, (, ), ''}).
The Logo Keyboard attenuates this problem and is, without doubts, one of
the advantages that the students most appreciated.

\section{Future work}
There is a great interest in the creation of educational interfaces
mainly with the Logo programming language.
In the near future, we intend to improve the Logo Keyboard with new
options, new configuration and new scanning
systems. We would like to continue to test the Logo Keyboard with
more individuals with motor impairments
including people with cerebral palsy.
We also intend to create interfaces that can help people with severe
motor impairments to control the movement
of a mouse quickly and easily in different Logo environments.
Our current goal is primarily focused on expanding the use of
computer programming among people with disabilities. We intend to work

\section{Conclusions}
This paper presented the Logo Keyboard, a system to be used by
individuals with physical disabilities to program in Logo.
The Logo Keyboard discussed in this paper is available for free use and can be
downloaded from \url{http://w3.ualg.pt/~snorte/LogoKeyboard}.
We hope that the Logo Keyboard will be useful to many people,
and that we can continue to explore new ways to aid
people with disabilities.  The observations and the studies with children
were very beneficial to understand
the needs and problems in the Logo learning environment. We expect that
Logo or other programming languages
may provide a good window for children to discover learning pedagogy and
problem solving methods.
Our work should also encourage researchers to explore new ways in which other
programming languages and essential techniques
can be used by users with disabilities.

\section*{Acknowledgments}
We sincerely thank the children and student testers in the school.
This work was sponsored by the Portuguese Foundation for Science
and Technology (FCT/MCES) under grant POCI/CED/62497/2004.

\bibliographystyle{plain}
\bibliography{references}

\end{spacing}
\end{document}